\documentclass[aps,prl,preprint,superscriptaddress,showpacs,twocolumn,10pt]{revtex4-1}
\usepackage{float}
\usepackage{graphicx}
\usepackage{epstopdf}
\usepackage{color}

\begin{document}

\title{Type-IV Pilus Deformation Can Explain Retraction Behavior}


\author{Ranajay Ghosh}
\affiliation{Department of Mechanical and Industrial Engineering, Northeastern University, Boston 02115 USA }
\author{Aloke Kumar}
\affiliation{Department of Mechanical Engineering, University of Alberta, Edmonton AB T6G 2G3,Canada  }
\author{Ashkan Vaziri}
\email{vaziri@coe.neu.edu}
\affiliation{Department of Mechanical and Industrial Engineering, Northeastern University, Boston 02115 USA }



\begin{abstract}
Polymeric filament like type IV Pilus (TFP) can transfer forces in excess of 100pN during their retraction before stalling, powering surface translocation(twitching). Single TFP level experiments have shown remarkable nonlinearity in the retraction behavior influenced by the external load as well as levels of PilT molecular motor protein. This includes reversal of motion near stall forces when the concentration of the PilT protein is lowered significantly. In order to explain this behavior, we analyze the coupling of TFP elasticity and interfacial behavior with PilT kinetics. We model retraction as reaction controlled and elongation as transport controlled process. The reaction rates vary with TFP deformation which is modeled as a compound elastic body consisting of multiple helical strands under axial load. Elongation is controlled by monomer transport which suffer entrapment due to excess PilT in the cell periplasm. Our analysis shows excellent agreement with a host of experimental observations and we present a possible biophysical relevance of model parameters through a mechano-chemical stall force map.
\end{abstract}
\maketitle
\begin{figure}
  \includegraphics [width=3.25 in,keepaspectratio=true]{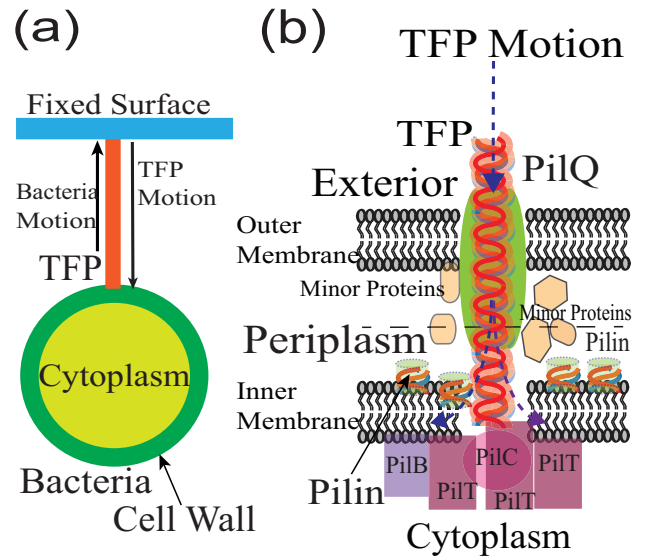}
  \caption{(a) Schematic illustration TFP spooling-in (retraction) action and the resultant bacterial motion resulting in twitching motility. (b)  a schematic illustration of the protein ensemble near the TFP base inside the cell wall which are responsible for the retraction process. The retraction/elongation process involves a large number of minor and primary proteins such as PilT, PilB molecular motors, PilC platform protein as well as the pore PilQ, all  spread across the periplasm of the cell. Pilins are stowed in the inner membrane after de-polymerization(retraction)and are subsequently recruited during polymerization (elongation). Dashed arrows indicate direction of motion~\cite{Skerker2001,Craig2008,Takhar2013}}\label{Fig.1}
\end{figure}
\begin{figure}
  \includegraphics [width=3.5 in,keepaspectratio=true]{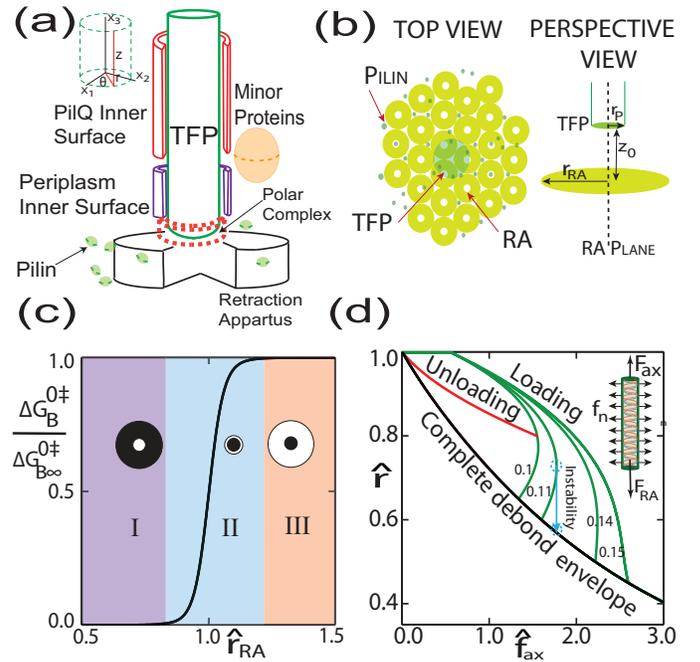}
  \caption{(a) A simplified reduction of the TFP processing bio-system into an axi-symmetric structure with a sliced view of TFP-protein/periplasm interfaces. The cylindrical retraction apparatus(RA) sits below the TFP base on the cytoplasmic part of the cell and the shallow  cylindrical ‘polar complex’ at the end of the TFP(shown in dotted red lines) is an electrostatic complex which is essential for recruiting pilins for elongation~\cite{Craig2008,Nivaskumar2014}(Insert:Cylindrical coordinates) (b) the top part of the RA (only PilT shown) forming the RA-plane is responsible for the binding regime of the retraction process and is assumed to be very closely packed with PilT units sitting close to the base of the TFP (c) binding energy at zero deformation as a function of size of the RA-plane indicating three distinct zones and a strongly saturating characteristic assuming a van-derWalls type binding. The x –axis is RA radius normalized by the pilus radius and y-axis is current binding energy normalized by that of an infinite plane. (Inserts: White circle indicates the size of RA plane and black the TFP cross section).(d)Normalized force-radius characteristic of TFP. The numbers on the loading curve (green) represent $\Delta G^0_C$(Insert: Free body diagram of loaded TFP,$F_{RA}$ is binding force due to RA)}\label{Fig.2}
\end{figure}

Elongation, adhesion and retraction of long polymeric nano-fiber called type-IV pilus (TFP) results in a form of bacterial surface translocation called twitching motility which causes complex colonization events such as virulence, biofilm formation and fruiting bodies~\cite{Mattick2002,Wall1999,Merz2000}. A host of proteins including molecular motors aid twitching motility through mechano-chemical processing of TFP, Fig.~\ref{Fig.1}(a) ~\cite{Bradley1980,Henrichsen1983,Wall1999,Wolfgang2000,Mattick2002,Burrows2005,Jin2011}. This highly repetitive processing consisting of rapid de-polymerization of TFP into pilins and the reverse - polymerization of the pilins into TFP near its base has been directly observed in \textit{Pseudomonas aerginosa}~\cite{Skerker2001}.
Among the ensemble of proteins responsible for TFP processing, the crucial role PilT protein~\cite{Satyshur2007,Misic2010}, a molecular motor, in aiding retraction was unambiguously isolated  and quantified in \textit{Neisseria gonorrhoeae}~\cite{Merz2000}. The \textit{in vivo} TFP retraction force-velocity characteristic of \textit{N. Gonorrhoeae} loaded using laser trapped micro bead showed constant retraction velocity at lower forces  which then decayed to a stable indefinite stall  as load was increased~\cite{Maier2002}. Interestingly, the retraction force-velocity characteristic was found to be nearly identical for mutants with differing concentration of PilT or periplasmic pilin. Later experiments on \textit{N. Gonorrhoeae} using similar set up showed that TFP retraction may even be reversed at stall fairly quickly into elongation for mutants with low concentration of PilT~\cite{Maier2004}.  More recent studies on \textit{N. Gonorrhoeae} have shown an yet undiscovered higher retraction velocity at lower forces for high PilT concentration mutants ~\cite{Clausen2009}. Thus, although the overall role of PilT protein in fostering TFP processing is beyond scrutiny, the exact interplay between force and PilT in altering force-retraction/elongation characteristic is intriguing thereby requiring assumptions beyond simple Arrhenius type kinetics~\cite{Clausen2009} or dynamics of a single Brownian motor or polymer ratchet mechanisms~\cite{Linden2006}.
In this work, we show that in contrast to the direct effect of force, the elasticity and geometry of the TFP together with its interfacial behavior when coupled with chemical kinetics play a key role in explaining the experimentally observed characteristics. This mechano-chemical paradigm which shows that retraction behavior is influenced by the characteristic of both the molecular motor and the TFP therefore point towards their \textit{coevolution} whose strong evidence for \textit{N. gonorrhoeae} has been reported in recent experiments~\cite{Biais2010}.
We first simplify the cell wall portion of TFP bio-system illustrated in Fig.~\ref{Fig.1}(b) into an equivalent homogenized axially loaded axi-symmetric cylindrical structure, Fig.~\ref{Fig.2} (a). The TFP is surrounded by a large protein PilQ spanning about half of periplasm, minor proteins as well as the periplasmic material itself~\cite{Burrows2012}. The TFP base may host a \textit{polar complex}(PC) which propels pilin recruitment through the charged end of growing TFP during elongation~\cite{Craig2008}. We simplify the arrangement of retraction proteins into a self-assembled axi-symmetric ensemble called retraction apparatus (RA) where motor proteins such as PilT play a leading role together with ancillary proteins such as PilC in TFP dis-assembly~\cite{Takhar2013,Craig2008}. PilT is a hollow cylinder which binds with the TFP at one end, excreting pilins at the other through large domain motion utilizing ATP hydrolysis~\cite{Satyshur2007,Misic2010}. This TFP consumption kinetics can be idealized as taking place in two steps via two distinct transition states (TS)-the first TS is part of the binding step which results in a metastable intermediate structure bound to the RA. The activation free energy for this reaction is mostly enthalpic in nature due to the binding field.
This meta-stable structure then disintegrates into pilins via another TS with the aid of PilT to mark the processing step which is likely entropic in nature due to polymer dismemberment and is independent of the binding field. The binding step determines the rate of forward and the processing step determines the rate of backward reaction. Note that this idealized kinetics subsumes the exact details of the still unclear molecular mechanism of this
transformation process involving a plethora of long and short range forces, interacting chemical species as well as thermally induced motion in highly complex condensed media through a unified reaction
coordinate. We idealize the binding as taking place between a sheet of binder surface  and the end cross section of the TFP with uniformly distributed binder sites,  Fig. ~\ref{Fig.2}(b). We compute the binding free energy $\Delta G_B^\ddagger$ (calculated per molecule of TFP material) by assuming vanderWalls(vdW) type interaction~\cite{Silverman2002} represented by an inverse sixth power pair- potential if the surfaces are sufficiently away from the steric repulsion regime (see Fig. 2(c)):
\begin{equation}
\Delta G_B^\ddagger=4\pi^2\sigma_{RA}A\int_0^{r_{RA}}\frac{r}{[z_0^2+(r-r_P)^2]^3}dr \label{Eq.1}
\end{equation}
here $r_{RA}$ is the radius of the retraction apparatus plane, $r_P$ is the current radius of the TFP, $\sigma_{RA}$ is areal density the binding site on the RA surface, $A$ is the vdW binding constant and $z_0$ is the inter-surface binding distance taken roughly equal to an average pilin characteristic length of $1nm$~\cite{Craig2006}. Eq.~\ref{Eq.1} can be re-written in terms of lengths normalized by TFP radius $r_P$ and plotted for various values of normalized RA radius $\hat r_{RA}=r_{RA}/r_P$, with $r_P=10nm$~\cite{Craig2006}, Fig.~\ref{Fig.2}(c).This plot exhibits a strong saturation characteristic, i.e. $\Delta G_B^\ddagger\approx\Delta G_{B\infty}^\ddagger, \hat r_{RA}\sim O(1)$  where $\Delta G_{B\infty}^\ddagger$ is the binding free energy of an infinite plane i.e. $\hat r_{RA}\rightarrow\infty$. Since the diameter of PilT is roughly of the order of the TFP itself, we conclude that increasing the concentration of PilT which would amount to increasing the size of RA will have little long term effect on retraction behavior as repeatedly confirmed in experiments~\cite{Maier2002,Maier2004}. Furthermore, evaluation of Eq.~\ref{Eq.1} in the infinite plane limit would yield:
\begin{equation}
\Delta G_B^\ddagger(\hat{r})\approx\Delta G_B^{0\ddagger}\hat{r}, \qquad \hat{r}=r_P/r_0\label{Eq.2}
\end{equation}
where $r_0$ is the undeformed TFP radius and the superscript 0 indicates the binding free energy under standard conditions of zero deformation, i.e. $r_P=r_0$.Thus the net areal mass production rate at TFP base for the retraction process $\dot{\mathcal{M}}_{ret}$ assuming unit chemical activity for  TFP and pilin material would be:
\begin{equation}
\dot{\mathcal{M}}_{ret}(\hat r)=k_0^+ e^{\Delta \hat G_B^\ddagger(\hat r)} - k^-, \qquad \Delta \hat G_B^\ddagger(\hat r)=\Delta G_B^\ddagger(\hat r)/k_BT\label{Eq.3}
\end{equation}
where $k_0^+$ is the rate constant without binding for the forward process, $k^-$ is the rate constant for the backward process, $k_B$ is the Boltzmann constant and $T$ is the temperature.  Note the TFP retraction velocity $v_{ret}=1/\rho_{TFP}\cdot\dot\mathcal{M}_{ret}$   where $\rho_{TFP}$ is TFP mass density.  In contrast to retraction process, elongation involves both polymerization and pilin transport towards the base of the TFP propelled by the electrostatic forces at the PC~\cite{Craig2008}, Fig.~\ref{Fig.2}(a). The PC however, must itself be stabilized for a steady pilin transport~\cite{Nivaskumar2014}. We propose that the stabilization is possible only when the net retraction rate has been diminished sufficiently. Once the incipient nucleus of the PC has stabilized, mass transport towards the TFP base commences resulting in the following flux- controlled elongation areal mass transport rate $(\dot\mathcal{M}_{el})$:
\begin{equation}
\dot\mathcal{M}_{el}=J_{flow}\cdot H[-\dot\mathcal{M}_{ret}]-(k_0^+e^{\Delta \hat G_B^\ddagger(\hat r)}-k^-)
\end{equation}\label{Eq.4}
\begin{figure}
  \includegraphics [width=3.25 in,keepaspectratio=true]{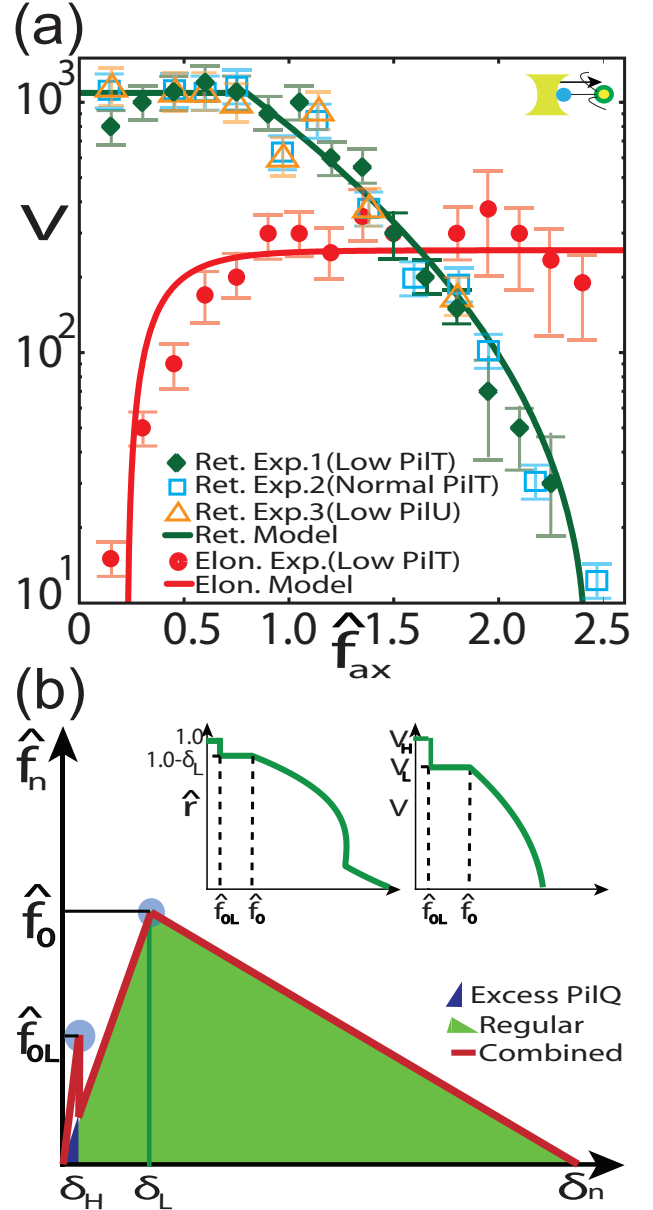}
  \caption{(a)Force-Bead velocity (in $nm/s$)comparison of the model with experiments~\cite{Maier2004}(Top Insert: Experimental setup)(b)traction-separation diagram showing progressive debonding $(\delta_i=1-r_i,\delta_H\ll\delta_L)$ and its effect on force-radius and force-velocity relationship which is now capable of reproducing the higher velocity mode of retraction $(V_H>V_L)$. $\hat f_{0L}$ is the lower cohesive strength(Inserts: Effect of progressive debonding on force-radius and force-velocity characteristics)}\label{Fig.3}
\end{figure}
where $H[\cdot]$ is the discrete Heaviside step function and $J_{flow}$ is out of RA plane transport flux at TFP base assumed independent of TFP diameter for the current work. Clearly, this thermodynamic framework depends on the TFP radius. In order to determine the evolution of TFP radius, we employ an elastic analysis of TFP deformation. To this end, first note that TFP is a multi-stranded helical structure~\cite{Craig2006,Craig2008}. Although some axial variation in geometry is possible, we approximate it as a regular n-start helical structure with a constant helix angle. A typical TFP with an outer radius of 10nm and inner radius of about 5nm~\cite{Craig2006}under about $150pN$ of peak stall force~\cite{Maier2004}  would be under a mean axial stress of less than 1MPa at a near stall loading rate of less than $20 nm/s$~\cite{Maier2004} implying negligible inelastic effects. Furthermore, electrostatic and thermal contribution to the strain energy are also neglected. In addition to the axial loading force, there are radial adhesive forces on the structure due to the volume surrounding the TFP as it runs through the enclosing PilQ, minor proteins as well as periplasmic gel, Fig. ~\ref{Fig.2}(a)~\cite{Burrows2012}. For simplicity, an average uniform adhesive traction is taken. All interfaces are assumed frictionless.  If the applied axial force is denoted by $F_{ax}$ and the adhesive traction as $f_n $, using Euler-Bernoulli theory, we arrive at the following force-radius relationship for the TFP assuming no unwinding (supplemental material):
\begin{equation}
\hat f_{ax}=L^2(\hat r)\left[ \hat f_n-\frac{n}{4}\frac{c_0^2}{\pi\hat r}\left\{M(\hat r)-\left(\frac{1}{1+\nu}\right)\left(1-\frac{\hat r}{L(\hat r)}\right)\right\}\right]\label{Eq.5}
\end{equation}
where $\nu,EI$ and $\alpha_0$ are respectively the Poisson's ratio, bending rigidity and the initial helix lead angle of the strands,  $r_0$ is undeformed radius of the TFP, $c_0=\cos \alpha_0$, $\hat f_n=\frac{f_n}{EI}r_0^4$, $\hat f_{ax}=\frac{F_{ax}}{\pi r_0^2}\frac{1}{EI}r_0^4\frac{n}{4}\frac{c_0^2}{1-c_0^2}$ , $L(\hat r)=\sqrt {\frac{1-\hat r^2c_0^2}{1-c_0^2}}$ and $M(\hat r)=c_0^2 \hat r \frac{\hat r-1}{1-\hat r^2c_0^2}$.The normalized adhesive traction $\hat f_n$ is modeled using the following traction-separation law~\cite{Park2011}:
\begin{equation}
\hat f_n(\hat r)=\Big\{^{\textrm{min}\{0,\hat f_0(1-\delta(\hat r)/\delta_n)\}, \qquad \textrm{load}}_{\textrm{min}\{0,\hat f_0(1-\delta_m/\delta_n)\delta(\hat r)/\delta_m\}, \qquad \textrm{unload}}
\end{equation}\label{Eq.6}
where $\delta(\hat r)=1-\hat r$, $\hat f_0$ is the normalized adhesive strength of the interface, $\delta_n$ is a dimensionless separation at complete failure and $\delta_m$ is the dimensionless separation at maximum load in case of partial failure. This relationship implicitly implies that the separation at which cohesive strength is reached, $\delta_c\ll1$ and thus non-dimensional interface cohesive free energy $\Delta \hat G_C=1/2\cdot\hat f_0\cdot\delta_n$ . Taking the geometrical properties of a typical \textit{N. gonorrhoeae} TFP, we have  $\alpha_0\approx  20^0, n=3$  ~\cite{Craig2006}. In addition, assuming a Poisson’s ratio of $\nu=0.45$ and $\hat f_0=0.55$, we generate the force-radius characteristic parameterized by $\Delta \hat G_C$, Fig.~\ref{Fig.2}(d). From here it is clear that lower $\Delta \hat G_C$ can result in material instabilities providing an instantaneous path for switchover from one branch to another thereby speeding the retraction-elongation switch as observed experimentally~\cite{Maier2004}. The portion of TFP external to the bacteria which is already under hydrostatic external pressure of the medium has been assumed pre-stretched by the time of debonding and thus does not contribute significantly to the retraction velocity. Although exact elastic parameters needed in the model have not been reported, we make indirect deductions. For instance, extension experiments on single TFP~\cite{Biais2010} have shown roughly a $40\%$ diametric reduction at forces of about $100pN$. Thus from Fig. 3(a),$\hat f_{ax}\sim67pN$. With these values, and using the following set of fitting parameters: $\nu=0.45,\hat f_0=0.8,\delta_n=0.45, k_0^+/\rho_{TFP}=5.53\times10^{-6} nm/s, k^-/\rho_{TFP}=0.553, J_{flow}/\rho_{TFP}=256.25 nm/s$ and $\Delta\hat G_B^{0\ddagger }=19.1$ in Eq. (3-5), we compare our model with single pilus elongation-retraction experiments~\cite{Maier2004} in Fig.~\ref{Fig.3}(a) (TFP geometrical properties have been kept as before) and find excellent agreement. Furthermore, in agreement with experiments~\cite{Maier2004}, retraction would resume as soon as laser trap is switched off since deformation vanishes causing instantaneous increase in radius and thus de-polymerization rate (Eq.~\ref{Eq.3}). Also, it has been found that only bacterial strains with low PilT concentration exhibit elongation but with indistinguishable retraction behavior when compared with normal or high PilT concentration strains~\cite{Maier2004}. This is a characteristic of our model where the elongation is exponentially attenuated by increasing levels of PilT in the inner membrane  due to increased pilin entrapment by PilT during transport. In the case where elongation is no longer possible due to a precipitous drop in pilin transport, the stall would represent a stable equilibrium. Although, purely concentration based diffusive transport has been ruled out since retraction rate was found to be indifferent to either the length of the retracted TFP or levels of pilin ~\cite{Maier2002}, any general transport process which suffers pilin entrapment due to PilT distribution in the periplasm would still exhibit this attenuation phenomena. The simplest model of uniform entrapment sites will lead to an exponential drop in mass transport rate with transporting distance~\cite{newton1982scattering} and thus, a higher level of PilT would also lead to much greater pilin entrapment leading to an eventual extinction of the incoming pilin mass flux beyond a threshold PilT concentration. Interestingly, areal density of entrapment sites would be directly related to only PilT units since they have a natural binding affinity for pilins and therefore, other co-expressed proteins (such as PilU) will have little effect on elongation; a claim which has already been confirmed by careful experiments~\cite{Maier2004}. Interestingly, since the transport step involving material transportation is slower than reaction the elongation process would exhibit pauses to allow for pilin buildup at TFP base, another experimentally observed hallmark ~\cite{Maier2004}. Recently, a higher far-from-stall retraction velocity (almost twice the average reported earlier)was observed at lower forces and high PilT concentration which abruptly switched to the widely observed lower retraction velocity as loading was increased~\cite{Clausen2009}.
\begin{figure}
  \includegraphics [width=3.25 in,keepaspectratio=true]{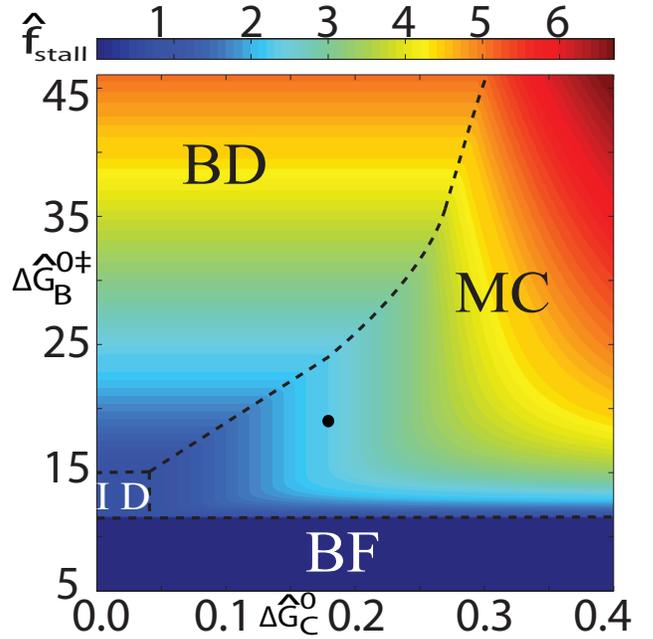}
  \caption{Stall-map indicating variation of normalized stall force with mechanical cohesive energy and normalized chemical binding activation free energy. Dashed lines show phase boundaries. The black circle shows the location corresponding to the experiments~\cite{Maier2004,Maier2002}. BF: Binding Failure, MC: Mechano-Chemical, BD: Binding Dominated and ID: Interface Dominated}\label{Fig.4}
\end{figure}
 We propose that excessive concentration of PilT causes an additional ring of PilTs to build up above the RA plane surrounding the TFP which although does not contribute to the retraction kinetics due to adverse steric position does provide another cohesive energy profile to the TFP. This profile is typically much weaker and more brittle than the existing profile due to poor contact and therefore alters the traction-separation law into a progressive one as shown in Fig.~\ref{Fig.3}(b). Therefore, instead of a single separation at which cohesive strength is reached i.e. $\delta_C$, there are two such separations: $\delta_H$ corresponding to the weaker PilT interface and $\delta_L\approx\delta_C$ correpsonding to the usual interface. Thus at $\delta_H$ TFP radius is $\hat r_H=1-\delta_H$ resulting in binding energy $\Delta\hat G_H^{0\ddagger }=\Delta\hat G^{0\ddagger}\hat r_H$. Similarly, at $\delta_L$, the binding energy is $\Delta\hat G^{0\ddagger}_L=\Delta\hat G^{0\ddagger}\hat r_L$. From Eq.~\ref{Eq.2} and Eq.~\ref{Eq.3}, we get $\delta_L-\delta_H\approx (1/\Delta\hat G_B^{0\ddagger})\ln{v_H/v_L}$ where $v_H$ and $v_L$ are respectively the retraction velocities (far from stall) of the higher and lower modes. Holding other model parameters constant, assuming $\Delta\hat G_H^{0\ddagger}\approx\Delta\hat G_B^{0\ddagger}$ and using experimental values~\cite{Clausen2009,Maier2004} we get $\delta_L-\delta_H\approx0.03$, implying $\hat r_H,\hat r_L\approx 1$, and thus $\hat L\rightarrow1,\hat M\rightarrow0$  in Eq.~\ref{Eq.5}. Therefore, this modification simply adds another step to the force-retraction curve at lower forces, Fig.~\ref{Eq.3}(b), thereby explaining the bimodal switching behavior. Note that due to inherently weak nature of this additional interface, this mode would be difficult to observe or sustain thus escaping detection in earlier 'spring loaded' experiments ~\cite{Clausen2009}.
We now generate a mechano-chemical stall plot in Fig.~\ref{Fig.4} which shows the landscape of normalized stall force variation depending upon $\Delta\hat G_C^0$ and $ \Delta\hat G_B^{0\ddagger}$ while other parameters are held constant from above. In this phase plot, at the bottom lies a binding failure region characterized by very low binding energy where retraction is decimated. As binding improves, we come across the next transitory interface dominated regime where binding energy is only large enough to be offset rapidly as soon as the interface fails, thereby constraining stall force to be near interface strength. As binding energy increases further, a binding dominated region emerges, where the stall force monotonically improves irrespective to the characteristic of the TFP interface. Bordering these regions lies the \textit{mechano-chemical} region where there is a complex interplay of the cohesive and the binding energy making it possible to arrive at a stall force through a relatively small variation of properties of both TFP interface and molecular motor.
Since higher levels of PilT can produce additional weaker interfaces as well, this  region provides maximum gains through PilT concentration changes. More specifically, in this region, poor alignment of PilT units due to excessive crowding which can otherwise reduce binding free energy and thus stall force may be mitigated automatically through additional cohesive energy. Thus the stall force which is an important parameter for survival and replication of these bacteria including biofilm formation and virulence~\cite{Merz2000} is much more robust in this mechano-chemical region. It is in this region that the experiments conducted on \textit{N. Gonorrhoeae}~\cite{Maier2002,Maier2004} lie and we believe this to be no coincidence as it boosts the evolutionatry adaptability of the organism. Furthermore, this region also provides a strong biophysical basis for \textit{coevolution} of both TFP properties as the underlying molecular motors, reported recently~\cite{Biais2010}. Note that although the experiments yielding the parameters were conducted on \textit{N. Gonorrhoeae}  TFP processing system is known to be extremely primitive and thus shows similar properties across a wide gamut of bacterial species thriving in widely different environmental landscape~\cite{Clausen2009}. Hence, conclusions drawn here are of broader biological significance.

RG, AV were supported by NSF CMMI Grant-1149750.

\bibliographystyle{unsrt}
\bibliography{Reference_Pilus}
\end{document}